\documentclass[aps, prd, preprintnumbers,twocolumn, showpacs, superscriptaddress, groupedaddress]{revtex4} 
\usepackage{graphicx}	
\usepackage{amssymb}
\usepackage{dcolumn}
\usepackage{color}
\usepackage{fancyhdr}
\usepackage{datetime2}
\usepackage{comment}
\usepackage{subfigure, rotating, bm, array}
\usepackage[pagebackref=false, colorlinks=true]{hyperref}
\hypersetup{
linkcolor=blue,     
citecolor=blue,     
urlcolor=blue}      
\begin{document}


\title{Relativistic orbits of S2 star in the presence of scalar field}

\author{Parth Bambhaniya}
\email{grcollapse@gmail.com}
\affiliation{International Center for Cosmology, Charusat University, Anand 388421, Gujarat, India}

\author{Ashok B. Joshi}
\email{gen.rel.joshi@gmail.com}
\affiliation{International Center for Cosmology, Charusat University, Anand 388421, Gujarat, India}

\author{Dipanjan Dey}
\email{deydipanjan7@gmail.com}
\affiliation{International Center for Cosmology, Charusat University, Anand 388421, Gujarat, India}

\author{Pankaj S. Joshi}
\email{psjcosmos@gmail.com}
\affiliation{International Center for Cosmology, Charusat University, Anand 388421, Gujarat, India}
\affiliation{International Centre for Space and Cosmology, Ahmedabad University, Ahmedabad, GUJ 380009, India}

\author{Arindam Mazumdar}
\email{arindam.mazumdar@iitkgp.ac.in}
\affiliation{Centre for Theoretical Studies, Indian Institute of Technology, Kharagpur, West Bengal-721302, India}

\author{Tomohiro Harada}
\email{harada@rikkyo.ac.jp}
\affiliation{Department of Physics, Rikkyo University, Toshima, Tokyo 171-8501, Japan}

\author{Ken-ichi Nakao}
\email{knakao@sci.osaka-cu.ac.jp}
\affiliation{Department of Physics, Graduate School of Science, Osaka Metropolitan University,
Sugimoto 3-3-138, Sumiyoshi, Osaka City 558-8585, Japan}
\affiliation{Nambu Yoichiro Institute of Theoretical and Experimental Physics, Osaka Metropolitan University,
Sugimoto 3-3-138, Sumiyoshi, Osaka City 558-8585, Japan}

\date{\today}
\preprint{RUP-22-20, AP-GR-183, NITEP 146}
\begin{abstract}
The general theory of relativity predicts the relativistic effect in the orbital motions of S-stars which are orbiting around our Milky-way galactic center. The post-Newtonian or higher-order approximated Schwarzschild black hole models have been used by GRAVITY and UCLA galactic center groups to carefully investigate the S2 star's periastron precession. In this paper, we investigate the scalar field effect on the orbital dynamics of  S2 star. Hence, we consider a spacetime, namely Janis-Newman-Winicour (JNW) spacetime which is seeded by a minimally coupled, mass-less scalar field. The novel feature of this spacetime is that one can retain the Schwarzschild spacetime from JNW spacetime considering zero scalar charge. We constrain the scalar charge of JNW spacetime by best fitting the astrometric data of S2 star using the Monte-Carlo-Markov-Chain (MCMC) technique assuming the charge to be positive. Our best-fitted result implies that similar to the Schwarzschild black hole spacetime, the JNW naked singularity spacetime with an appropriate scalar charge also offers a satisfactory fitting to the 
observed data for S2 star. Therefore, the JNW naked singularity could be a contender for explaining the nature of Sgr A* through the orbital motions of the S2 star.

\vspace{.6cm}
$\boldsymbol{key words}$ : Black hole, Naked singularity, Periastron precession, Milky way, S2 star.
\end{abstract}
\maketitle

\section{Introduction} 
Recently, the Event Horizon Telescope collaboration has announced a major breakthrough in the imaging of an ultra-compact object at the centre of our galaxy \cite{EventHorizonTelescope:2022xnr,EventHorizonTelescope:2022vjs,EventHorizonTelescope:2022wok,EventHorizonTelescope:2022exc,EventHorizonTelescope:2022urf,EventHorizonTelescope:2022xqj}. A bright emission ring around a core brightness depression in VLBI horizon-scale images of Sgr A*, with the latter linked to the shadow of black hole. The shadow boundary of the Sgr A* marks the visual image of the photon region and differentiates capture orbits from scattering orbits on the plane of a distant observer. The radius of the bright ring can be used as an approximation for the black hole shadow radius under specific conditions and after proper calibration, with little reliance on the details of the surrounding accretion flux. While there is strong evidence that there is a high concentration of mass in the center of our Milky Way galaxy, the question of whether or not it is a black hole is still open. They have considered various alternatives such as naked singularities and regular black holes. They favorably acknowledge that the naked singularity with a photon sphere Joshi-Malafarina-Narayan (JMN-1) naked singularity could be the best black hole mimicker \cite{EventHorizonTelescope:2022xqj}. The central object and its nature remain mysterious. This is because just like a black hole case, the JMN-1 naked singularity would create a similar shadow, and therefore it is very difficult to distinguish between the two. Therefore, in this paper, we study the relativistic orbits of stars that are orbiting around our own galactic center. 

Near the center Sgr A* of our Milky-way galaxy, many stars are hovering around with very high speed. Due to the presence of the central massive object of mass around $4\times 10^6 M_{\odot}$, these stars can move at $\frac{1}{60}$ the speed of light and they can have highly eccentric orbits. These stars are known as `S'-stars. Since they are very close to the galactic center, there exists a possibility that they can show up some general relativistic effects. However, it is very difficult to follow the dynamics of those stars, since they are far away from us ($\approx 25,000$ light years). The highly sensitive infra-red instruments namely GRAVITY, SINFONI, and NACO in the European Southern Observatory (ESO) are capable of tracking the trajectory of the `S'-stars. Recently, they have released the 23 years of astrometric data of the `S2' star \cite{Do:2019txf,GRAVITY:2018ofz,GRAVITY:2020gka} which is one of the important star of the `S' star family. As it is mentioned above, general relativistic effects can be seen in the dynamics of the `S2' star, and therefore, its trajectory can give us information about the spacetime around the Sgr A*. In \cite{DellaMonica:2021xcf,DellaMonica:2021fdr,deMartino:2021daj}, authors investigate the nature of Sgr A* with the geodesic motion of S-stars in Scalar-Tensor-Vector Gravity, f(R ) gravity and with  ultralight bosons.
 
It is a general belief that the spacetime of the central supermassive object (Sgr A*) is a vacuum solution of Einstein field equations (e.g., Schwarzschild solution, Kerr solution, etc.). However, from the observational results, it can be understood that the center of a galaxy is surrounded by highly concentrated matter. Therefore, a vacuum solution is more unlikely to be present around a galactic center. Therefore, one could consider a matter distribution near the central region and investigate the corresponding physical signature. There are several papers where non-vacuum spacetimes are considered and the different physical signatures (e.g. shadow, timelike orbits, accretion disk, etc.) of the same are investigated \cite{Bambhaniya:2019pbr,Dey:2019fpv,Joshi:2019rdo,Joshi:2020tlq,Bambhaniya:2021ybs,Bambhaniya:2021ugr,Solanki:2021mkt,Saurabh:2022jjv,Vagnozzi:2022moj}.

In this paper, we investigate the scalar field effect on the orbital dynamics of the `S2' star. A scalar field is the simplest constitute of matter and at the beginning, one may model the matter distribution around the galactic center using the scalar field. Here, we do not discuss anything about the particle physics model of the scalar field. It may be interpreted by some beyond the standard model (BSM) of particle physics which is not the scope of the present paper. Here, we are interested in constraining the scalar charge by best-fitting the astrometric data of `S2' with the theoretical prediction using the MCMC technique. In order to do that, we consider JNW spacetime which is the minimally coupled, mass-less scalar field solution of Einstein equations. The JNW spacetime possesses a naked singularity at the center. One can retain the Schwarzschild spacetime from the JNW spacetime by considering zero scalar charge. However, a small but non-zero scalar field charge can drastically change the causal structure of the Schwarzschild spacetime. The best-fitted value of the scalar charge comes out to be non-zero which may be interpreted as the existence of non-vacuum spacetime seeded by a scalar field around the galactic center. 

We organize this paper as follows:
In Sec. (\ref{II}), we discuss the properties of timelike geodesics in JNW spacetime. In Sec. (\ref{III}), we discuss the orbital parameters of the real and apparent orbits. In Sec. (\ref{IV}), we constrain the orbital parameters' space of JNW spacetime by best-fitting the astrometric data of the `S2' star with the theoretical prediction using the MCMC technique. In Sec. (\ref{V}), we discuss our results with concluding remarks.

\section{Timelike geodesics in JNW spacetime}
\label{II}
The JNW spacetime is a mass-less, minimally coupled scalar field solution of Einstein field equations, which is spherically symmetric and static spacetime and it is given by \cite{Virbhadra:1997ie,Janis:1968zz}
\begin{widetext}
\begin{eqnarray}
 ds^2 &=& -\left(1-\frac{b}{r}\right)^n c^2dt^2 + \left(1-\frac{b}{r}\right)^{-n}dr^2 + r^2\left(1-\frac{b}{r}\right)^{1-n}d\Omega^2\,\, ,
 \label{JNWmetric}
\end{eqnarray}
\end{widetext}
where $d\Omega^2=(d\theta^2 + \sin^2\theta d\phi^2)$, $b=2\sqrt{\left(\frac{GM}{c^2}\right)^2+q^2}$ and $ n=\frac{2GM}{c^2b}$, here
$G$ and $M$ are the gravitational constant and ADM mass of the spacetime and $q$ is the charge of the mass-less scalar field. It can be seen from the above equation that the JNW spacetime becomes Schwarzschild spacetime for zero scalar charge ($q$).   As the JNW is a spherically symmetric, static spacetime, the conserved energy and angular momentum per unit rest mass can be written as,

\begin{eqnarray}
\frac{e}{c^2} = \left(1-\frac{b}{r}\right)^n\left(\frac{dt}{d\tau}\right),\,\, 
\end{eqnarray}

\begin{eqnarray}
L = r^2\left(1-\frac{b}{r}\right)^{1-n}\left(\frac{d\phi}{d\tau}\right),\,\, 
\end{eqnarray}
using these conserved quantities in equation (\ref{JNWmetric}), we can define the total relativistic energy as, 
\begin{equation}
    E_n=\frac{\frac{e^2}{c^4}-1}{2}=\frac{1}{2}\left[\frac{1}{c^2}\left(\frac{dr}{d\tau}\right)^2+V_{eff}(r)\right],
\end{equation}
\noindent
where, $V_{eff}(r)$ is the effective potential of the JNW naked singularity spacetime. One can derive the following timelike orbit equation for the JNW spacetime, 
\begin{widetext}
\begin{eqnarray}
\frac{d^2u}{d\phi^2}+u-\frac{3bu^2}{2}+\frac{e^2b}{c^2L^2}(1-n)(1-bu)^{1-2n}-\frac{c^2b}{2L^2}(2-n)(1-bu)^{1-n}=0\,\, ,
\label{orbitgen}
\end{eqnarray}
\end{widetext}
where $u=\frac1r$.
\begin{figure}
    \centering
    \includegraphics[width=7.8cm]{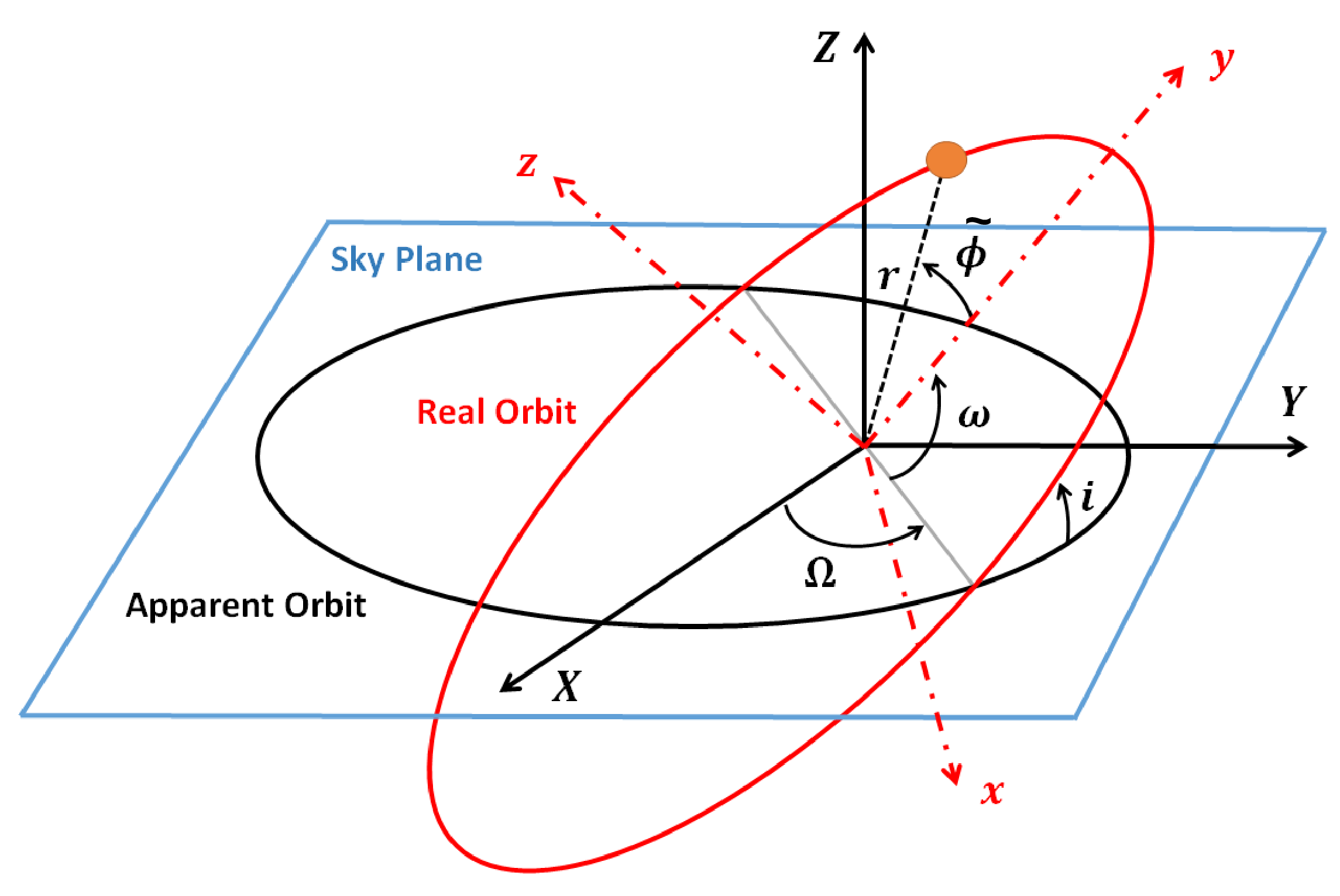}
    \caption{Real orbit projection into the sky plane \cite{Becerra-Vergara:2020xoj}. The axes begin at Sgr A* (the focus of the ellipse). The following diagram depicts the orbital parameters: $\tilde{\phi}=\phi-\pi/2$, where $\phi$ is the azimuth angle of the spherical coordinate system associated with the $x, y, z$ Cartesian coordinates, i.e. the true anomaly for an elliptic motion in the $x-y$ plane, $i$ is the angle of inclination between the real orbit and the sky plane, $\Omega$ is the ascending node angle, and $\omega$ is the pericenter argument.
It is worth mentioning that the vector going from the solar system to the galactic centre defines the coordinate system's Z-axis.}
    \label{orbitproject}
\end{figure}
In \cite{Bambhaniya:2019pbr,Dey:2019fpv}, the properties of timelike orbits in the JNW spacetime are elaborately discussed and also compared with the timelike orbits in Schwarzschild spacetime. The important difference which is coming out from the analysis is that the perihelion precession of bound timelike orbits in the JNW spacetime, can be negative, i.e., the direction of particle motion is opposite to the direction of precession. This unique characteristic of timelike orbit is forbidden in Schwarzschild spacetime. In \cite{Igata:2022rcm,Igata:2022nkt}, it is shown that the presence of matter may be responsible for the negative precession. 
Therefore, negative precession is very much important in the context of the trajectories of `S' stars around the Sgr A*. In the next section, we use the above orbit equation to predict the possible trajectory of `S2' star, and using the astrometric data of that star, we constrain the parameters' space of the JNW spacetime.

\begin{figure*}
    \centering
    \includegraphics[width=13cm]{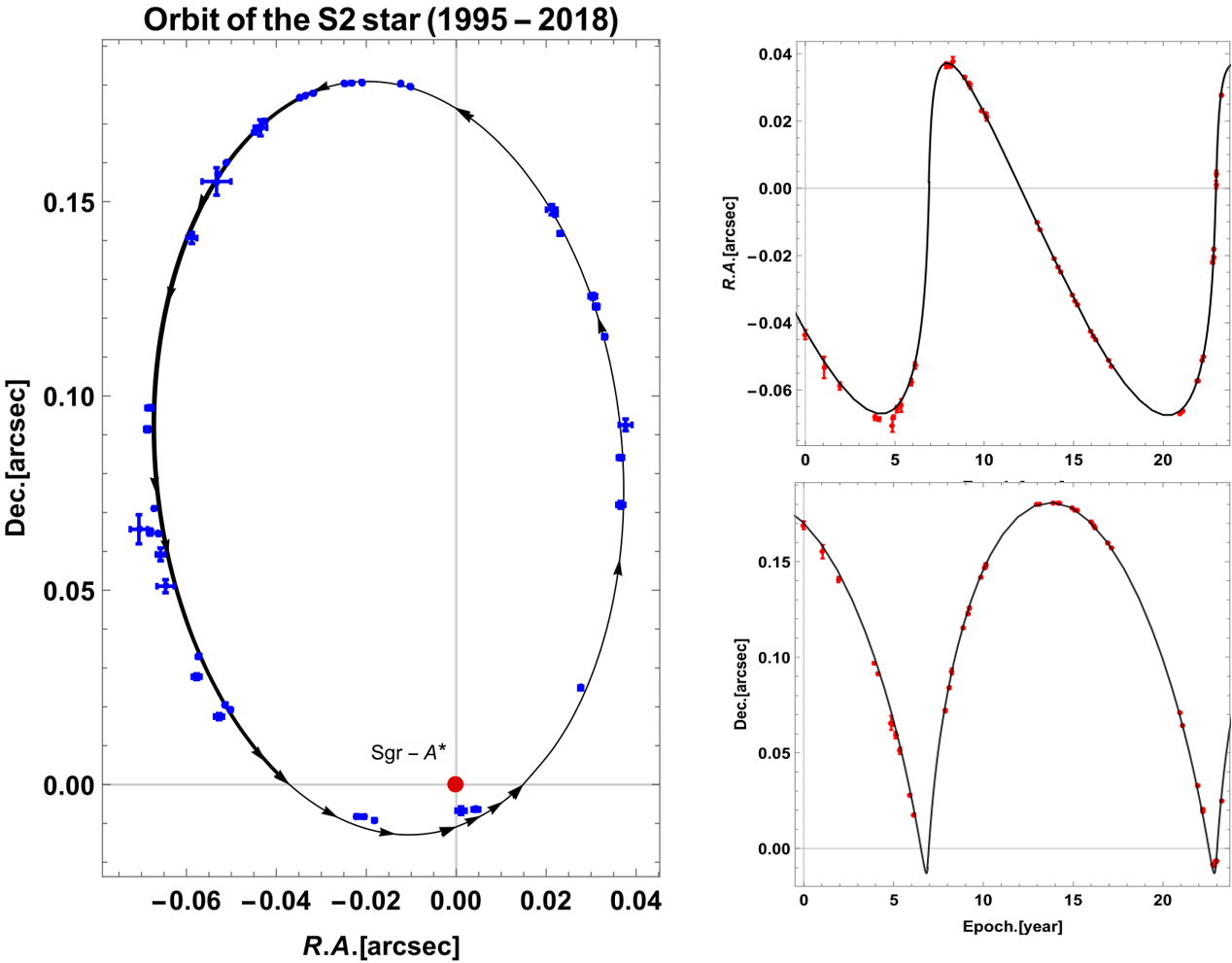}
    \caption{Left: Best fitting orbit of the JNW model (black) along with the observed astrometric position (Blue) of the S2 star from 1995 to 2018. The red dot shows the position of the Sgr A* at (0,0). Right: R.A. (right ascension $\alpha$ on top) and Dec. (declination $\delta$ on bottom) offset of S2 star with the orbital period.}
    \label{fig:orbit_JNW}
\end{figure*}

\section{Orbital parameters of the real and apparent orbits}
\label{III}
It is necessary to transform the real orbit into the apparent orbit, since the observed astrometric data given onto the plane of the sky. We define the position and velocity components of the real orbit in Cartesian coordinates as $x, y, z$ and $v_x, v_y, v_z$ respectively. As in our case, $\theta=\pi/2$, we can obtain position and velocity of the real orbit by the transformation from spherical coordinates to Cartesian coordinates as
\begin{equation}
    x=r\cos \phi, \; \; \;
    y=r\sin\phi, \; \; \;
    z=0,
    \label{position1}
\end{equation}
and the corresponding three velocities transform as,
\begin{equation}
    v_x=v_r\cos \phi-rv_{\phi}\sin\phi, \; \; \;
    v_y=v_r\sin\phi+rv_{\phi}\cos\phi, \; \; \;
    v_z=0,
    \label{velocity1}
\end{equation}
where, $v_r=dr/dt$ and $v_{\phi}=d\phi/dt$ are three velocity components and the corresponding four velocity can be written as $u^r=v_r u^0$ and $u^{\phi}=v_{\phi} u^0$. Now, to fit the real orbit with the astrometric observational data, we must project the real orbit on the apparent plane of the sky as shown in fig. (\ref{orbitproject}). The observed astrometric positions $X_{obs}$ and $Y_{obs}$ of the star in Cartesian coordinates are defined by the observed angular positions, right ascension $(\alpha)$ and declination $(\delta)$ \cite{Becerra-Vergara:2020xoj}.

\begin{equation}
    X_{obs}=r_d (\alpha-\alpha_{SgrA^*}),\; \; \;
    Y_{obs}=r_d (\delta-\delta_{SgrA^*}),\; \; 
\end{equation}
\begin{table}
    \centering
    \vspace{0.05cm}
    \begin{tabular}{l c c}
        \hline\hline
    &       &         \\
{\bf Parameter}      &                                 {\bf JNW (95\% limits)}   \\
\hline
\hline
\\
{\boldmath$L^2         $} $(pc^2(km/s)^2)$ &  $4.44216^{+0.00075}_{-0.00075}$ \\
\\
{\boldmath$\log{E_n}          $} $(km/s)^2$ &  $10.95422391^{+0.00000034}_{-0.00000033}$ \\
\\
{\boldmath$t_{\rm ini}          $} (year) &  $1.199^{+0.038}_{-0.040}   $ \\
\\
{\boldmath$\log{M}         $} $(M\odot)$ & $6.666^{+0.010}_{-0.012}   $ \\
\\
{\boldmath$\log{q}          $} $(M\odot)$ & $-7.46^{+0.58}_{-0.57}     $ \\
\\
{\boldmath$\theta_{\rm inc}        $} (radian) & $2.316^{+0.025}_{-0.025}   $ \\
\\
{\boldmath$\Omega         $}  (radian)& $4.017^{+0.035}_{-0.033}   $ \\
\\
{\boldmath$\omega         $} (radian) & $1.199^{+0.029}_{-0.029}   $ \\
\\
{Distance (parsec), \boldmath$r_d          $} & $8169^{+34}_{-39}          $ \\
\\
{Time period (yr), \boldmath$T         $} & $16.1379          $ \\\\
{Minimal \boldmath$\chi^2         $} & $4.71          $ \\
\hline
    \end{tabular}
    \caption{Estimated best-fit values of the parameters for the JNW metric.}
    \label{table2}
\end{table}
where $r_d$ is the distance between the Sgr A* and the earth. Note that the center of the co-ordinate system represent the position of Sgr A*. The positions $X, Y, Z$ of the apparent orbit can be obtained from the real orbit positions $x$ and $y$ by using classic Thiele-Innes constants with the same notation given in \cite{Becerra-Vergara:2020xoj} as
\begin{equation}
    X=xB+yG, \; \; \;
    Y=xA+yF, \; \; \;
    Z=xC+yH, \; \;
    \label{position2}
\end{equation}
 and the corresponding velocity components of the apparent orbit are,
 \begin{equation}
    V_X=v_xB+v_yG, \; \; \;
    V_Y=v_xA+v_yF, \; \; \;
    V_Z=v_xC+v_yH, \; \;
    \label{velocity2}
\end{equation}
where,
\begin{equation}
    A=\cos\Omega\cos\omega-\sin\Omega\sin\omega\cos i,
\end{equation}
\begin{equation}
    B=\sin\Omega\cos\omega+\cos\Omega\sin\omega\cos i,
\end{equation}
\begin{equation}
    C=\sin\omega\sin i,
\end{equation}
\begin{equation}
    F=-\cos\Omega\sin\omega-\sin\Omega\cos\omega\cos i,
\end{equation}
\begin{equation}
    G=-\sin\Omega\sin\omega+\cos\Omega\cos\omega\cos i,
\end{equation}
\begin{equation}
    H=\cos\omega\sin i,
\end{equation}
where the osculating orbital elements $\Omega$, $i$, and $\omega$ are the ascending node angle, inclination angle, and the argument of pericenter, respectively. Now, the fully general relativistic solution of the above equation (\ref{orbitgen}) gives the $r(\phi)$, which we have to transform into Cartesian coordinates using the relation given in (\ref{position1}). The apparent orbital coordinates $X, Y$ can be obtained from the real orbital positions $x, y$ by using the transformation (\ref{position2}). Now, we can fit the observed apparent orbital data with the apparent orbital plane, which will give us the actual nature of the orbital shape.

\section{MCMC analysis}\label{IV}
The MCMC analysis performed in the paper for the astrometric data of S2 star is based upon the Metropolis-Hastings algorithm \cite{Hastings:1970aa}. The likelihood function used in the analysis for symmetric error is as follows,
\begin{equation}
-\log {\mathcal L} \propto \sum_i \left[\left(X_i - \bar{X}_i\over \sigma^X_i\right)^2 + \left(Y_i - \bar{Y}_i\over \sigma^Y_i\right)^2\right].
\end{equation}
Here $X_i$ and $Y_i$ represent the observed co-ordinates of the S2 star in its orbit and $\bar{X}_i$ and $\bar{Y}_i$ are theoretically calculated values. The errors of the observation in the $X$ and $Y$ direction are $\sigma^X$ and $\sigma^Y$ correspondingly. To
implement the MCMC analysis, the value of $q$ is always
put within the open bounded interval $(0, q_0)$, where $q_0$ is
taken to some large positive value. This means that the
present analysis excludes the possibility of $q = 0$, i.e., the
Schwarzschild spacetime, all along.

Details about the data set: 46 number of data points are considered here from the astrometric positions of the S2 star. Source of the orbital data is adopted from the supplement material of the paper \cite{Do:2019txf}. For the MCMC analysis we used Gaussian priors. The details of the priors are given in the table~(\ref{tab2}).

\begin{table}
    \vspace{0.2cm}
    \begin{tabular}{|l | c |c|}
\hline
    &       &\\        Parameter & Mean & 1-$\sigma$ \\
     \hline
    &       &\\      {\boldmath$L^2          $} ($pc^2(km/s)^2$) & $4.44$ & $0.04$

\\

{\boldmath$\log{E_n}          $} & $10.90$ & $0.01$\\

{\boldmath$t_{\rm ini}          $} (yr) & $1.22$  & $0.01$ \\

{\boldmath$\log{M}          $} & $6.56$ & $0.05$ \\

{\boldmath$\log{q}          $} & $-8.00$ & $0.005$ \\

{\boldmath$\theta_{\rm inc}          $} (radian) & $2.30$ & $0.02$ \\

{\boldmath$\Omega          $} (radian) & $4.00$ & $0.04$ \\

{\boldmath$\omega          $} (radian) & $1.20   $ & $0.01$\\

{Distance, \boldmath$r_d          $} (parsec) & $8200$ & $10$\\
\hline
\end{tabular}
\caption{Details of the Gaussian priors of different parameters.}
\label{tab2}
\end{table}
\section{Conclusion}
\label{V}
\begin{figure*}
    \centering
    \includegraphics[width=\linewidth]{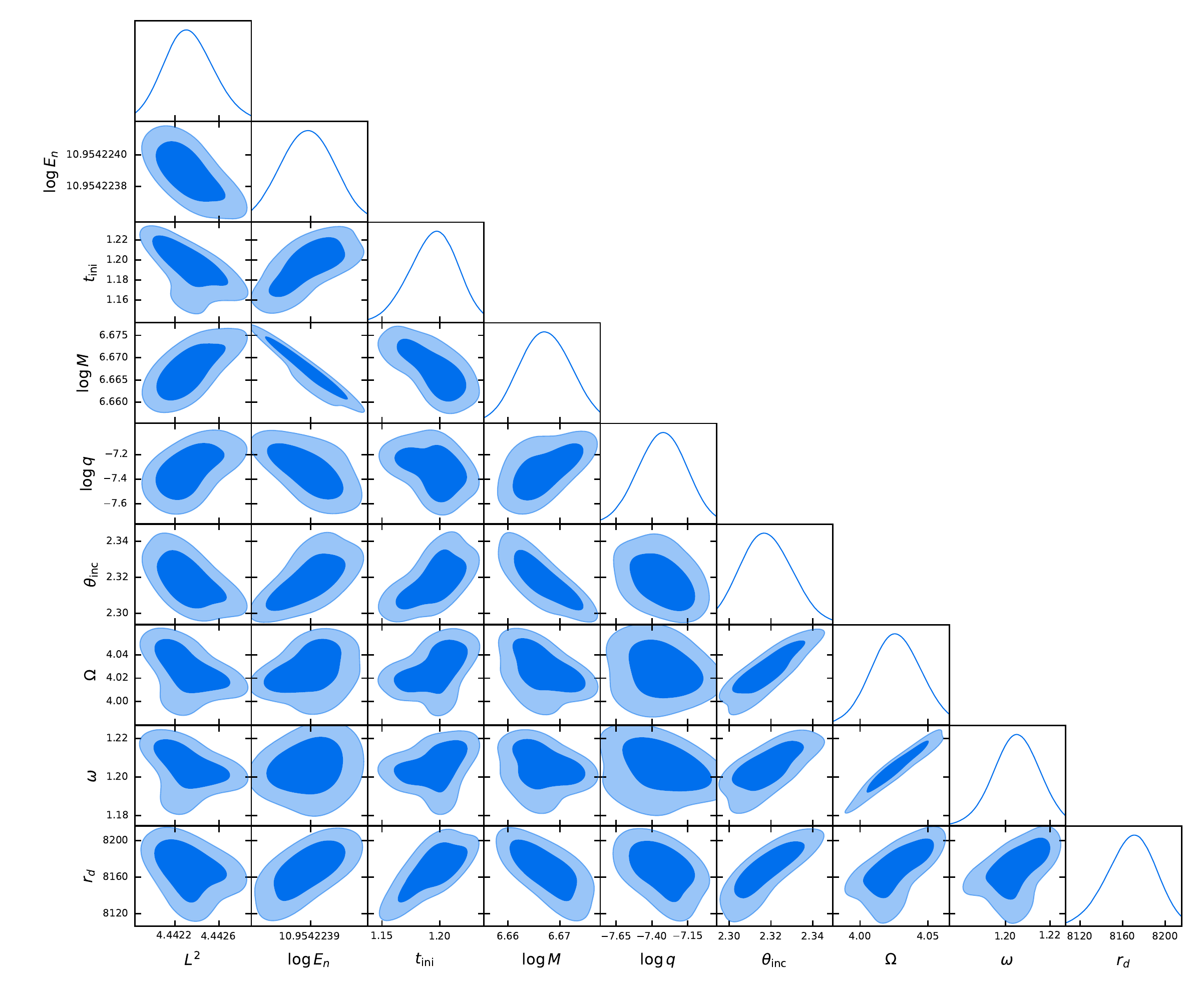}
    \caption{1-$\sigma$ and 2-$\sigma$ bestfit regions and the posterior distributions of the parameters for the JNW metric and Sgr A* derived using MCMC. Lowest $\chi^2$ value obtained is 4.71.}
    \label{fig:param_JNW}
\end{figure*}

In this paper, we have derived the fully relativistic orbit equation for the JNW spacetime, which is a second-order non-linear differential equation. We solve this equation numerically since it is difficult to solve analytically. To obtain the best fitting orbital parameters of the JNW model, we use the astrometric data of S2 star from \cite{Do:2019txf}. We estimate the best fitting parameters for the JNW metric using the MCMC technique under the assumption of the positive scalar charge and obtain the lowest $\chi^2$ value is 4.71 (see the figures (\ref{fig:orbit_JNW}) and (\ref{fig:param_JNW})). Here, we predict the nature of the Sgr A* using the available observed astrometric data of the S2 star. Our results show that like Schwarzschild black hole, the JNW naked singularity could be a possible candidate for the compact object Sgr A* at our Milky Way galaxy center. 

\acknowledgments
TH is very grateful to H. Saida for fruitful discussions and helpful comments. This work was partially supported by JSPS KAKENHI Grant Nos. JP19K03876, JP19H01895, JP20H05853 (TH) and JP21K03557 (KN).

\end{document}